\documentclass[11pt,cospar]{article}
\usepackage{cospar}
\usepackage{url}

\usepackage{graphicx}

\title{Real time localization of Gamma Ray Bursts with INTEGRAL}

\author{
S. Mereghetti\address{IASF/CNR Milano, v. Bassini 15, Milano, I 20133, Italy},
D. G\"{o}tz$^{1}$
and
J. Borkowski\address{INTEGRAL Science Data Center, Versoix, Switzerland}
}

\begin{document}

\maketitle

\medskip

\begin{center}

34th COSPAR Sci. Assembly, Houston, 10-19 october 2002 

\end{center}

\begin{abstract}

The INTEGRAL satellite has been successfully launched in October 2002
and has recently started its operational phase.
The INTEGRAL Burst Alert System (IBAS) will distribute
in real time  the coordinates of the GRBs detected with INTEGRAL.
After a brief introduction on the INTEGRAL instruments, we describe the main IBAS
characteristics and report on the initial results.
During the initial performance and verification phase
of the INTEGRAL mission, which lasted about two months,
two GRBs have been localized with accuracy of $\sim$2-4 arcmin.
These observations have allowed us to validate the IBAS software, which is now expected
to provide  quick (few seconds delay) and precise (few arcmin)
localization for $\sim$10-15 GRBs per year.

\end{abstract}

\section*{THE INTEGRAL MISSION}
\vspace{3pt}

INTEGRAL is a satellite of the European Space Agency devoted to high-resolution
imaging and spectroscopy in the hard X--ray / soft $\gamma$-ray energy range
(Winkler et al 1999).
After a very  successful  launch from the Baikonur Cosmodrome in Kazachstan by a Russian
Proton rocket on October 17, 2002, INTEGRAL has been injected  into a three day period
orbit with perigee of $\sim$9,000 km and  apogee of $\sim$153,000 km.
Thanks to this highly eccentric orbit, the satellite spends most of the
time outside the Earth radiation belts, where it can carry out long, uninterrupted
observations.

The INTEGRAL spacecraft   carries
two main    instruments operating in the $\sim$15 keV -- 10 MeV
energy range. They are optimized respectively for spectroscopy (SPI, Vedrenne et al. 1999)
and imaging (IBIS, Ubertini et al. 1999) performances. Both are able to obtain images
of the $\gamma$-ray sky using the coded aperture technique.

SPI  is based on an array
of 19 Germanium detectors cooled to 85 K to provide an energy resolution of
$\sim$2 keV at 1 MeV. It has a wide field of view (diameter $\sim$30$^{\circ}$) and
an angular resolution of $\sim$2$^{\circ}$, adequate for a detailed
mapping of the diffuse line emission in the Galactic plane.

Since IBIS is the instrument more relevant for the
accurate localization of GRBs, we will describe it in more detail.
IBIS  has been optimized for the imaging performances
by means of a   coded mask,  made of
square tungsten elements with a thickness of 15 mm, placed   at a distance
of $\sim$3 m from the detection plane. The mask, that accounts for about
one third of the total IBIS weight ($\sim$650 kg),  absorbs
$\sim$70\% of the photons at 1.5 MeV, in correspondence of its closed elements, while
the mask mechanical support retains an adequate transparency at low energy  for
the open elements. The good contrast of the coded aperture pattern is in fact directly
related to the instrument sensitivity. The size of the mask elements yields an
angular resolution of 12 arcmin, unprecedented at the
energies covered by IBIS,  and   appropriate to resolve sources in
crowded regions, such as the Galactic Bulge.
The IBIS field of view is very large:  $\sim29^{\circ}\times29^{\circ}$ (at zero response),
and with a uniform sensitivity within its  central part  ($\sim10^{\circ}\times10^{\circ}$).

In order to
effectively cover an extended energy range, the IBIS detection plane
is composed of two layers with
different characteristics and based on different technologies. The top layer,
ISGRI (Lebrun et al. 2001),   consists of an array of 128$\times$128 Cadmium
Telluride (CdTe) square pixels (4 mm $\times$ 4 mm each).
It covers the energy range from $\sim$15 keV up to a few hundreds keV
with a   total sensitive area of  $\sim$2600 cm$^{2}$.
The CdTe is a semiconductor that can be used at ambient temperature providing a good
energy resolution ($\sim$7\% at 100 keV) without the need of complex cooling
systems. ISGRI is used  in  ''photon'' mode, i.e.  the
complete information of each detected event (position, energy  and arrival time)
is recorded and transmitted to the ground.

Due to the CdTe thickness of 2 mm, the ISGRI detection plane becomes
almost transparent for photons above $\sim$200 keV,
but  such photons will be efficiently stopped by PICsIT, the second layer of the
instrument. PICsIT  is based on an array of 64$\times$64 scintillators of
Caesium Iodide doped with Thallium (CsI(Tl)). Each element is a small bar of
CsI(Tl) with dimensions 8.6$\times$8.6$\times$30 mm$^{3}$ coupled with a photodiode.
Although the two IBIS layers are in principle operating as   independent imaging
systems, their data will be used in a combined analysis to better study
the spectral properties of the observed sources over the whole
energy range. In addition, a fraction of the detected photons will undergo a
Compton interaction in ISGRI, before being detected by PICsIT. For such
''Compton'' events, the energy and position measurements provided by the two
layers (which are separated by $\sim$10 cm), will give some constraints on the
possible arrival directions, allowing to increase the signal to noise ratio by
appropriate selections aimed at excluding directions not modulated by the mask
aperture or not compatible with the position of the source of interest.

The two main INTEGRAL instruments, SPI and IBIS, are complemented by an X-ray
monitor (JEM-X, 4-35 keV) and an optical camera (OMC) operating in the V band.
All the INTEGRAL instruments are co-aligned and will be used simultaneously
to provide a broad energy coverage of the targets in the central part of the
IBIS and SPI field of view.

\section*{INTEGRAL AND GRBs}
\vspace{3pt}

Although the INTEGRAL satellite  was not specifically  conceived as a mission
devoted to GRBs studies, it was soon realized that the excellent imaging
performances of IBIS could offer the possibility of rapid localization
of the events observed by chance in its large field of view.
It was therefore proposed to implement a ''burst alert system'' in order to allow rapid
multi-wavelength  follow-ups. Such a
proposal was boosted by the   results obtained with the
\textit{BeppoSAX} satellite that clearly
demonstrated the excellent capabilities of  coded
mask instruments to quickly detect and locate GRBs.

Accurate estimates based on the LogN-LogS relations for GRBs derived from BATSE led
to a predicted rate of at least one GRB per month in the IBIS field of view
(Mereghetti et al. 2001a),
for which positions at the arcmin level can be obtained.
Note that the source location accuracy and the angular resolution, though related,
are two different concepts.
Although the IBIS angular resolution (i.e. the width of the point response
function) is 12 arcmin,  sources  can be located with a much better
accuracy, which is a function of the
signal to noise ratio. For instance, a source with a signal to
noise ratio of 30 can be located by ISGRI with an accuracy of 30$''$ (90\%
c.l.).

Exploiting the continuous telemetry downlink of the INTEGRAL satellite,
it was decided to carry out the search for GRBs
on ground,  at the INTEGRAL Science Data Center (ISDC, Courvoisier et al. 1999).
The ISGRI data normally  reach the ISDC
within a few seconds from their detection on board  the satellite.
When a  GRB is detected, its coordinates are automatically distributed
via internet to the subscribed clients.



\section*{THE INTEGRAL BURST ALERT SYSTEM}
\vspace{3pt}

The INTEGRAL Burst Alert System (IBAS) is the automatic software running at
the ISDC for the rapid detection and localzation of GRBs (Mereghetti et al. 2001b).
In order to optimize the response time, IBAS  is
independent of the main data processing pipeline of the ISDC.

Fig. 1 gives an overview of the  IBAS software architecture.
The telemetry, received at the ESA Mission Operation Center
in Darmstadt, is  continuously  transmitted to the ISDC on a 128 kbs dedicated
line, where the
Near Real Time Data Receipt
subsystem extracts the relevant telemetry packets
and, after some basics checks,  feeds them into the IBAS subsystem.
IBAS basically consists of several \textit{Detector Programs} and the
\textsl{Ibasalertd Program}.
All these programs run in parallel. The \textit{Detector Programs} have the task to
trigger on possible GRBs and to perform preliminary checks to filter out,
as much as possible, spurious events. The \textit{Ibasalertd} program has the task to
further filter   the triggers, by combining in an appropriate way the information
from the different \textit{Detector} programs, and to eventually send out the GRB Alerts.
All the IBAS processes are multi-thread applications written in C or C++.
They run as daemon processes, which means that they do not perform any
terminal input/output and run in background. IBAS processes perform several
subtasks in parallel, each one handled by a separate thread.
With some unavoidable exception, the subtasks are independent and do not
block each other.

This architecture allows us to use in parallel different methods for the GRB detection,
as well as to run several instances of the same \textit{Detector Program} with different
parameter settings (e.g. timescales, energy cuts, etc...) in order to increase the
sensitivity for  GRBs with different properties.
Currently two algorithms to find and locate GRBs in data from ISGRI are in use,
plus one program to detect GRBs seen by the anticoincidence shield of SPI
(in this case no directional information is available).
Other \textit{Detector} programs based on data from JEM-X and SPI are under development.

\begin{center}
\includegraphics[width=105mm]{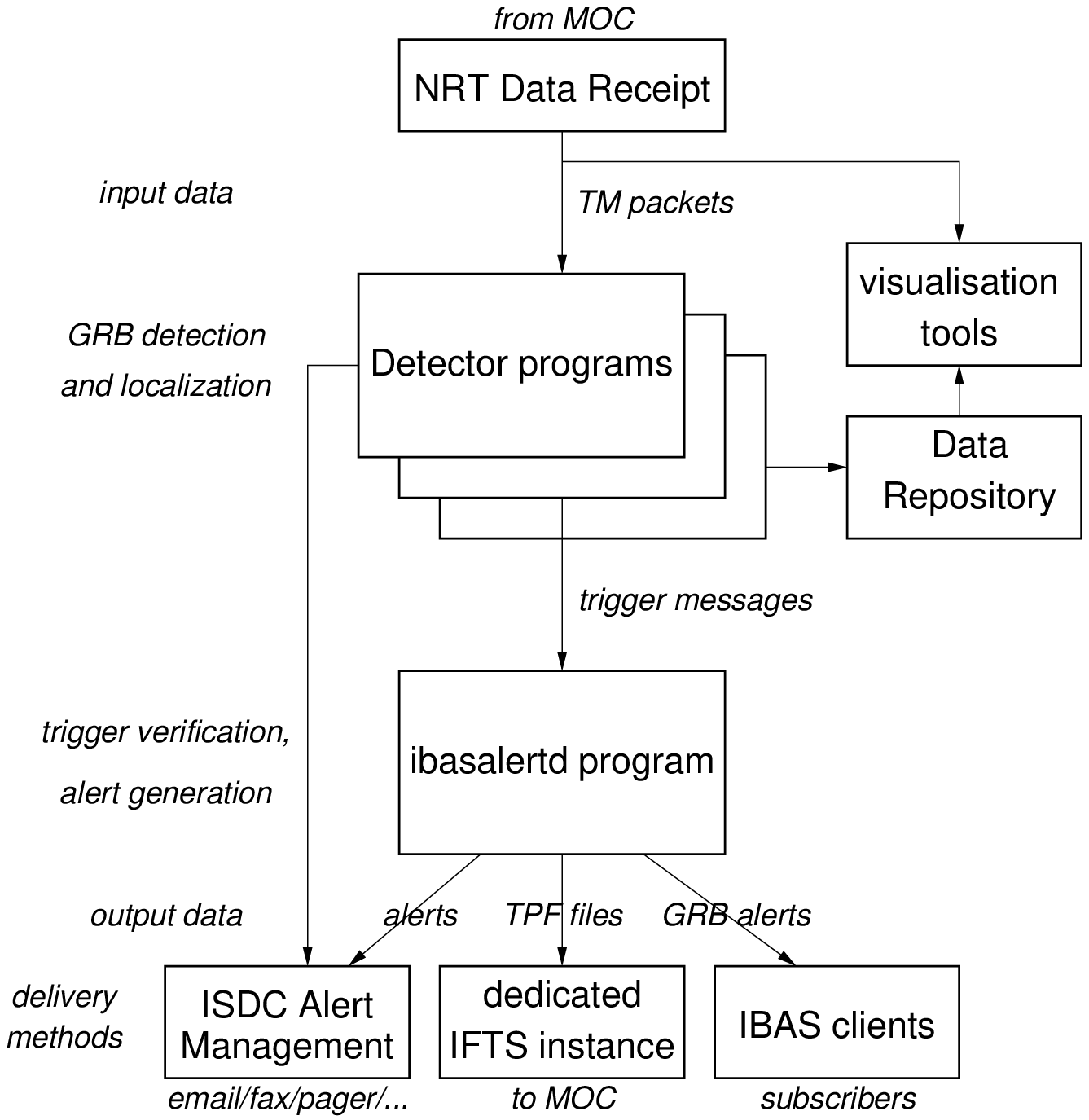}
\end{center}

{\sf Fig. 1.
Main components of the IBAS software.
}
\vspace{8mm}

The \textit{Ibasalertd Program} is designed  to combine the information
it receives from the \textit{Detector Programs}. The basic adopted logic  is that triggers
with high statistical significance are considered as valid GRBs even if they
come from a single \textit{Detector Program}, while triggers of lower significance require
an independent confirmation from different data.
The details of the  logic of trigger confirmation
can be defined in a very flexible way by means of a set of parameters
involving significance of detection,
tolerance for positional and temporal coincidence, etc...

If the event is genuine,  the \textit{Ibasalertd Program} determines the best available
satellite attitude information  to derive the GRB sky position, which
is then delivered to the IBAS Clients by means of \textsl{Alert Packets}.
In addition, if the GRB is located in the central 5$^{\circ}$$\times$5$^{\circ}$
of the IBIS field of view, which is covered by the OMC,
the information  is  transmitted to the MOC,
where an  appropriate telecommand is automatically generated and sent to the
satellite. This will reconfigure the observing mode of the OMC in order to
place a  CCD window  on the position of the  GRB.

\subsection*{Distribution of the IBAS Alert Packets}

IBAS \textit{Alert Packets} with the GRB information are sent via Internet,
using the UDP transport protocol.
We have developed a \textit{Client Software} that allows the users to receive the
\textit{Alert Packets} and to easily use their content in the software commanding
their telescopes. The IBAS \textit{Client Software}, written in standard C language
and tested on Sun Solaris, Linux and MacOSX, and its documentation can be
downloaded from the ISDC web pages.
The package contains one library and two simple
programs (one to check the connection with the computer on which IBAS is
running and one to print on the screen the content of the \textit{Alert Packets}).
Each packet is 400 bytes long, and consists of several fields, the
format and content of which is explained in detail in the Client software
documentation.

As described above, the IBAS GRB detection and alert distribution is mostly based on
automatic processes. An interactive analysis, to confirm the event and
to derive the most accurate GRB position, will generally be performed a few
hours after the automatic delivery of the first alert message(s) with the
preliminary coordinates.
In designing an automatic GRB alert system, one is faced with the trade-off
between the requirement to react in the shortest possible time and the
desire to get the most accurate results (e.g. reality of the event, dimension
of error region, etc...).
Considering that the main users of the IBAS Alerts will be robotic telescopes
with relatively large fields of view, specifically devoted to GRB afterglow searches,
we have privileged the fast alert time requirement.
This necessarily implies that some of the IBAS alerts might
subsequently be found not related to real GRB events. Users not interested
in the fastest reaction time can decide to receive only particular
alert types.
Five types of \textit{Alert Packets}  have been defined:

\vspace{3pt}

\underline{Packet type 1 : POINTDIR.}

Many automated telescopes can exploit the \textit{a priori} knowledge
of the INTEGRAL pointing direction (e.g. to reduce the slew time in case
of a GRB alert, to obtain reference images of the pre-GRB sky, to monitor
the counterparts of INTEGRAL targets). IBAS will send
a new POINTDIR packet each time a slew to a new pointing direction begins.
The POINTDIR packet contains the pointing direction expected at the end
of the slew and the expected time of slew end.

\vspace{3pt}
\underline{Packet type 2 : SPIACS.}

Alert packets of type SPIACS will be
sent after positive triggers
detected with the SPI Anti Coincidence Shield. No position information
will be available, unless the same GRB also triggers some of the imaging
instruments (in such a case the position will come through other packet
types). SPIACS packets contain the name of the file with the GRB light curve
that is immediately  available on the public ftp server of the ISDC.
These light curves can be used in conjunction with those of other satellites
to locate the GRB with the time delay triangulation method.

\vspace{3pt}
\underline{Packet type 3 : WAKEUP.}

Packets of type WAKEUP are generated only
once for each GRB and
only if the GRB position information is available. They contain the preliminary
position (R.A., Dec. J2000) derived for the GRB. These are the alerts with
the shortest time delay (and consequently the highest chance of not being
due to a real GRB event). Any (potential) GRB will generate only one WAKEUP
packet. IBAS will then distribute more refined information as soon as it
becomes available using packets of type 4.

\vspace{3pt}
\underline{Packet type 4 : REFINED.}

These Packets provide refined information on a GRB event.
IBAS will automatically generate a new REFINED packet only if
the new information available has significantly smaller uncertainties and/or
supersedes the one sent with previous packets. Zero to several REFINED packets
can be generated for a given GRB.

\vspace{3pt}
\underline{Packet type 5 : OFFLINE.}

These Packets are generated manually after an
interactive analysis of the
data performed off-line. The typical time delay can be from
one to a few hours (or even longer in some cases).
The OFFLINE alert packet represents the
final confirmation (or rejection) of a GRB and contains the most accurate
GRB properties distributed by IBAS.

\section*{FIRST RESULTS}
\vspace{3pt}

The first two months after the launch of INTEGRAL have been devoted to the
initial set up and activation of the instruments and to the
Performance and Verification phase.
During this period, the IBAS system has been operated each time suitable data
were available (but with the automatic delivery of \textit{Alert Packets} to external clients
switched off).
This allowed us to further test the system and to perform minor
changes to the software in order to better adapt it to the in-flight performances
of the instruments.
At the time of writing (December 2002) INTEGRAL has already detected two
GRBs in the field of view of IBIS: GRB021125 and GRB021219.
The public distribution of the IBAS Alert Packets will begin in January 2003.

GRB021125 (Bazzano and Paizis 2002) occurred during an observation
in which most of the the available
satellite telemetry was allocated to PICsIT for calibration purposes.
Thus only a limited fraction of the events recorded
with ISGRI could be transmitted to the ground.
Unfortunately, due to the resulting gaps in the ISGRI data, the IBAS \textit{Detector
Programs} were in idle mode at the time of the burst and did not trigger in real time.
The IBAS imaging programs were used in the off-line interactive analysis to
locate the burst (see Figure 2). Further analysis resulted in a localization  with
an uncertainty of only 2 arcmin (Gros and Produit 2002),
consistent with that independently derived with the IPN (Hurley et al. 2002a).
After this event, we introduced some changes in the IBAS \textit{Detector Programs} in order to
be able to detect  GRBs also during observations with limited ISGRI telemetry allocation
(note, however, that such an instrument configuration is not supposed to occur
often during routine operations).

The second GRB in the IBIS field of view, GRB021219 (Mereghetti et al. 2002),
was found and correctly localized by the IBAS software in real time.
The position  derived by the IBAS software within $\sim$10 seconds of the burst occurrence
had an accuracy of $\sim$20 arcmin,  largely dominated by systematic uncertainties.
In less than four hours these could be significantly reduced resulting in a 4 arcmin
error box (G\"{o}tz et al. 2002), which was subsequently confirmed with
the IPN (Hurley et al. 2002b), as shown in Figure 3.

\begin{center}
\includegraphics[width=8cm]{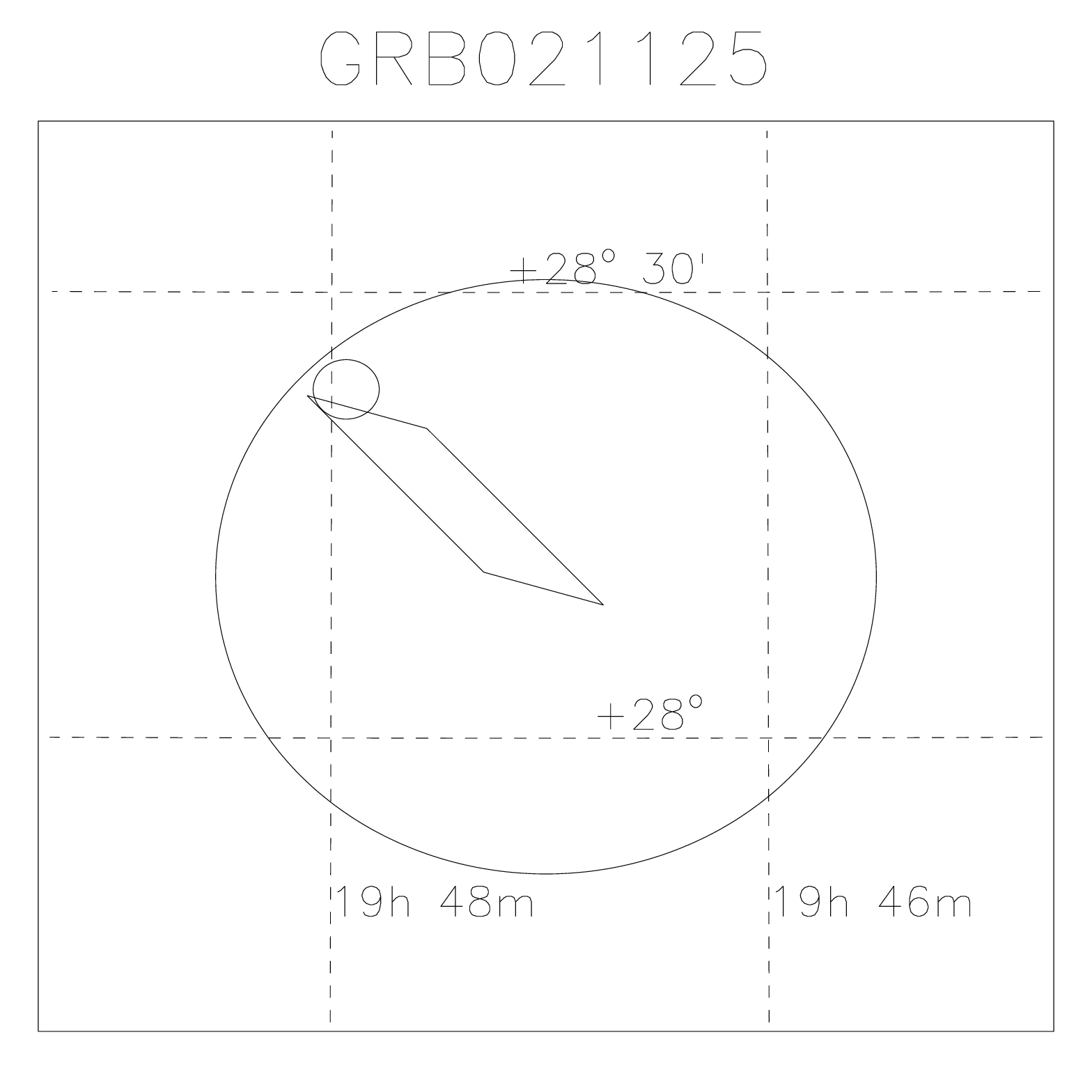}
\end{center}

{\sf Fig. 2.
\textsl{Localization of GRB021125. The large circle (20$'$ radius) is the first
error region obtained with the IBAS programs, while the small one
is the refined position (2$'$ radius) obtained after a preliminary
boresight correction (Gros and Produit 2002).
The parallelogram is the IPN error box (Hurley et al. 2002a).
}
\vspace{8mm}

\begin{center}
\includegraphics[width=8cm]{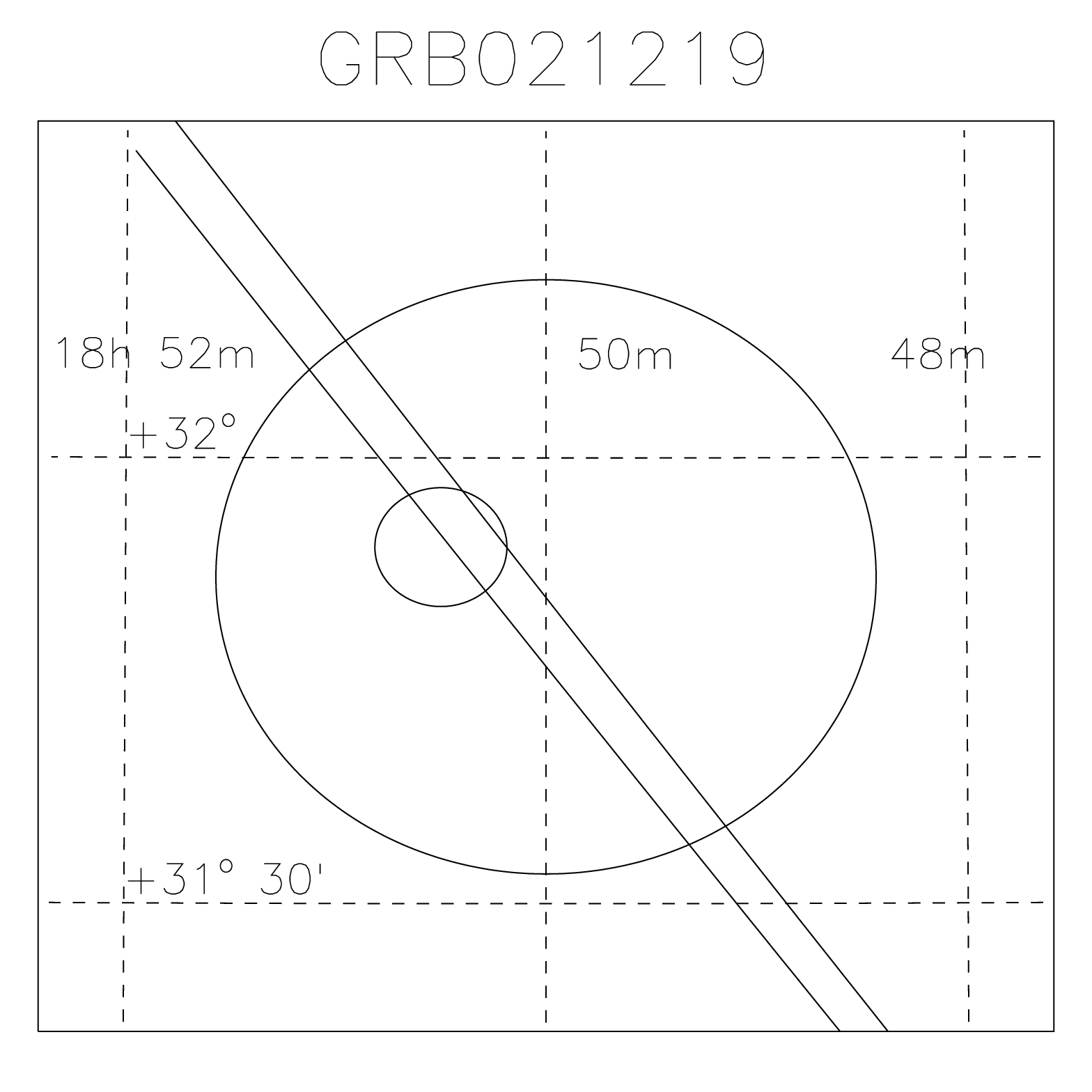}
\end{center}

{\sf Fig. 3.
Localization of GRB021219.
The first IBAS error circle has a radius of 20$'$ (Mereghetti et al. 2002),
The final localization obtained with an off-line analysis, shown by the
small circle  (4$'$ radius, G\"{o}tz et al. 2002) is consistent with the IPN annulus
(Hurley et al. 2002b).
}
\vspace{8mm}



\section*{ACKNOWLEDGEMENTS}
\vspace{3pt}

We thank Davide Cremonesi and Don Jennings for their   contribution
to the initial development of the IBAS software.
SM and DG acknowledge the  support of the Italian Space Agency.
JB acknowledges support from the Polish State Committee for Scientific Research
(grant number 2P03C00619p02).
We thank the ISWT for permission of using PV data for the testing of IBAS.

\end{document}